\documentclass[conference]{IEEEtran}
\IEEEoverridecommandlockouts

\usepackage{cite}
\usepackage{amsmath,amssymb,amsfonts}
\usepackage{algorithmic}
\usepackage{graphicx}
\usepackage{textcomp}
\usepackage{xcolor}
\usepackage{multirow}
\usepackage[table,xcdraw]{xcolor}
\def\BibTeX{{\rm B\kern-.05em{\sc i\kern-.025em b}\kern-.08em
    T\kern-.1667em\lower.7ex\hbox{E}\kern-.125emX}}

\begin{document}
\title{
    A Core-Structure-Based Automated Deobfuscation Tool for Commercial Virtualization Obfuscation Analysis\\
}
\author{
    \IEEEauthorblockN{Wanju Kim}
    \IEEEauthorblockA{\textit{Dept. of Computer Sci. \& Eng.}  \\
    \textit{Chungnam National University}\\
    Daejeon, Republic of Korea \\
    201902669@o.cnu.ac.kr}
    \and
    \IEEEauthorblockN{Seoksu Lee}
    \IEEEauthorblockA{\textit{Dept. of Computer Sci. \& Eng.}  \\
    \textit{Chungnam National University}\\
    Daejeon, Republic of Korea \\
    troy.doubles@o.cnu.ac.kr}
    \and
    \IEEEauthorblockN{Eun-Sun Cho}
    \IEEEauthorblockA{\textit{Dept. of Computer Sci. \& Eng.}  \\
    \textit{Chungnam National University}\\
    Daejeon, Republic of Korea \\
    eschough@cnu.ac.kr}
}
\maketitle
\begin{abstract}
    Virtualization obfuscation is a more powerful obfuscation technique compared to other obfuscation methods, and as it is increasingly being applied to malware, it demands significant effort and time from analysts. 
    This study analyzes virtualization obfuscation and proposes a tool called VMPredator that automatically extracts semantic units. 
    The proposed tool performs various analyses including memory analysis and trace analysis, while minimizing dependency on the specific internal structure of virtual machines in order to handle diverse forms of virtualization obfuscation that existing tools are unable to process. 
    Experimental results demonstrate that the length of obfuscated programs was reduced by approximately 85\%, and it was verified through validation that small-scale programs were fully restored to semantics identical to the original.
\end{abstract}
\begin{IEEEkeywords}
    component, formatting, style, styling, insert.
\end{IEEEkeywords}
\section{Introduction}
Malware authors employ a variety of obfuscation techniques to evade detection and hinder analysis. 
In particular, virtualization obfuscation -- an obfuscation technique provided by commercial obfuscation tools -- transforms original code into user-defined virtual instructions and executes them on a custom virtual machine, making it one of the most powerful code protection mechanisms in existence. 
This approach completely conceals the structure of the original code and renders static analysis virtually impossible, thereby impeding the rapid analysis of malware by security professionals. 
Furthermore, deobfuscating virtualization obfuscation requires a high level of expertise and considerable time.

Accordingly, efforts have been made to automate the deobfuscation of virtualization obfuscation. 
However, if an analysis tool models a specific VM structure in too much detail, it becomes excessively dependent on that particular implementation.
Conversely, without any assumptions about the obfuscation structure, the amount of recoverable information becomes too limited for automation to be practically useful. 
In particular, commercial obfuscation tools may employ variant forms such as threaded dispatch to eliminate well-known structural cues like the decode-dispatch loop, which further reduces the generalizability of approaches that rely on fixed CFG patterns.

This paper seeks a practical middle ground between these two extremes by defining key elements -- VM entry, VM exit, and handlers -- as semantic anchor points, thereby securing meaningful structural information necessary for automated analysis without being tied to any specific obfuscation structure. 
Based on this approach, an automated tool is implemented and evaluated on commercial VMProtect samples, confirming that it demonstrates higher effectiveness than existing publicly available automated techniques. 
The contributions of this paper are as follows:

\begin{itemize}
    \item We propose VMPredator, a fully automated analysis framework for commercial virtualization obfuscation that does not rely on overly specific structural assumptions.
    \item We design and utilize a minimal structural assumption that employs only three core semantic elements: VM entry, VM exit, and handlers.
    \item We demonstrate applicability not only to decode-dispatch loops but also to cases where VM structural characteristics have been eliminated, such as threaded dispatch variants.
    \item We logically verify the success of semantic unit recovery for each extracted unit.
    \item We present experimental results showing higher effectiveness compared to existing publicly available automated analysis tools against the commercial obfuscation tool VMProtect.
\end{itemize}

\section{Backgrounds}
    \subsection{Virtualization Obfuscation and Related Work}\label{AA}
Virtualization obfuscation is an obfuscation technique that transforms original instructions into user-defined virtual instructions for execution.
Rather than executing the original code directly, it interprets and executes the code on an arbitrary virtual machine (VM), thereby providing strong resistance against both static and dynamic analysis.

A common method for implementing a virtual machine is the decode-dispatch loop approach. 
In this scheme, the decode stage decodes the next opcode and operands to be executed and dispatches them to the appropriate virtual instruction handler, repeating this process until the virtual machine terminates. 
Since this structure resembles a switch-case construct inside a while loop, it is possible to identify the pattern and analyze the structure of the embedded virtual machine.

Prior work has leveraged this structural insight to deobfuscate virtual bytecode\cite{b7}\cite{b8}. 
However, modern commercial virtualization obfuscation tools have introduced a threaded approach to conceal these characteristics. 
In the threaded approach, each virtual instruction handler is assigned its own unique decode-dispatch routine. 
For example, after a virtual instruction handler performs the operation corresponding to that handler, it executes the handler-specific decode-dispatch routine to compute the address of the next handler to be dispatched. 
This effectively hides the typical while-loop switch-case pattern of conventional virtual machines, thereby neutralizing existing analysis methods.
Deobfuscation techniques that focus on handling diverse virtualization obfuscation structures extract semantics without relying on the virtualization structure at all. 
it is therefore expected that additional mechanisms such as separate heuristics are employed to achieve the reported performance\cite{b3}.

    \subsection{Virtual Instruction Handler}
In this study, we analyzed the virtual machine structure generated by VMProtect 3.5, the most widely used commercial virtualization obfuscation toot\cite{b1}. 
Since the advanced edition of VMProtect adopts the threaded approach, no centralized decode-dispatch routine exists, rendering conventional virtualization deobfuscation methods that analyze routine structure ineffective.
Therefore, this study identifies the fundamental operational semantics of each virtual instruction handler and tracks only those operations. 
To this end, we define the most fundamental semantic unit -- the "virtual instruction handler" -- as follows:

[Definition 1] 
A Virtual Instruction Handler is an operational unit that retrieves operands from the virtual context -- consisting of a virtual stack and virtual registers -- performs the operation corresponding to the virtual instruction, and writes the result back to the virtual context.

\begin{figure}[b]
    \centering
    \includegraphics[width=0.5\linewidth]{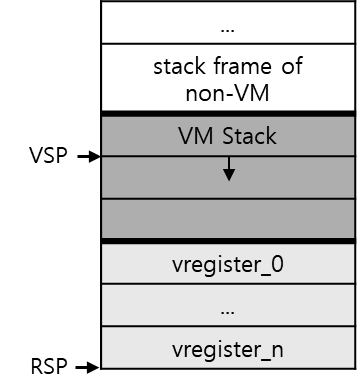} 
    \caption{Structure of the Virtual Stack and Virtual Registers in the Virtual Machine Used by VMProtect} 
    \label{fig:f1}
\end{figure}

To distinguish virtual instruction handlers, we first analyzed the virtual context, which represents the virtual machine's stack and registers. 
VMProtect reuses the existing stack frame to serve as the virtual context for storing virtual machine operation results. 
As illustrated in Figure \ref{fig:f1}, when the virtual machine is initialized, the rsp register is decremented to allocate space, which is then partitioned into a virtual stack and virtual registers. 
The virtual stack is managed by the VM's VSP (Virtual Stack Pointer) and grows in the direction of decreasing addresses, identical to the x86 stack structure.

Furthermore, a virtual instruction handler in a typical threaded virtual machine operates according to the following structure: \\
(1) To perform an instruction operation or value transfer, values are fetched from the virtual context or the operands of the virtual instruction and stored in physical registers. \\
(2) If values are fetched from the virtual instruction's operands, a decoding process is applied to those operands. \\
(3) The operation corresponding to the handler is executed. Finally, \\
(4) the operation result held in the physical registers is written back to the virtual context. 

Once the result is stored, the decode-dispatch routine is executed to compute the address of the next handler. 
After the address of the next handler is calculated into the jump register, an indirect jump instruction is used to transfer control to the next handler.

\section{Main Idea: VMPredator}
Rather than modeling all VM structures, this paper extracts a minimal set of semantic anchors -- core structural elements that remain relatively stable even when commercial obfuscation tools conceal their structural characteristics -- and uses them as the basis for analysis.

\begin{figure}[b]
    \centering
    \includegraphics[width=0.5\linewidth]{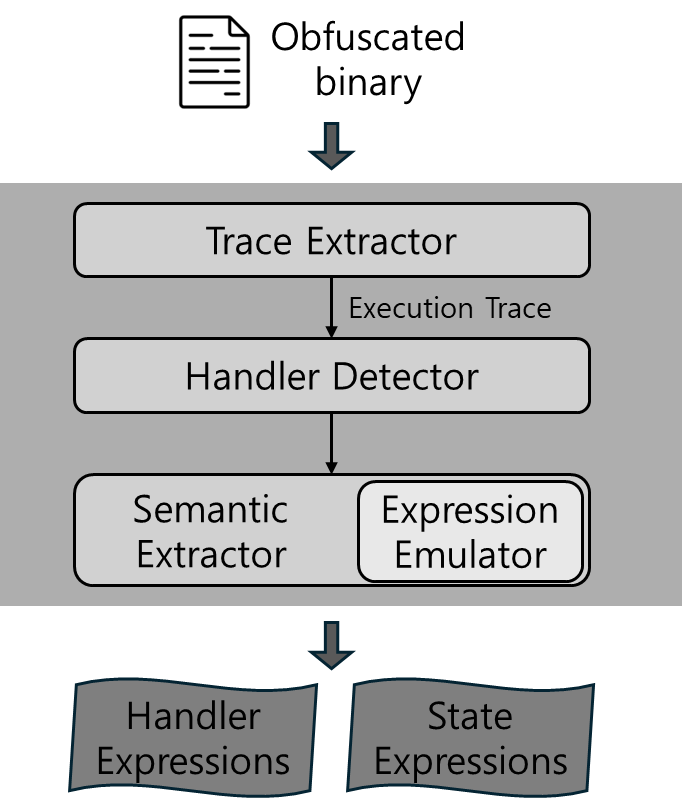} 
    \caption{Architecture of VMPredator} 
    \label{fig:f2}
\end{figure}

[Definition 2] 
A semantic anchor is a core structural element that is invariantly preserved across diverse variants of virtualization structures.

This study defines semantic anchors as the triple ⟨VM Entry, VM Exit, set of virtual instruction handlers⟩ and, based on this analysis, proposes VMPredator -- a tool that automatically extracts operational sequences from the execution trace of an executable obfuscated by a commercial obfuscation tool. 
As illustrated in Figure \ref{fig:f2}, the proposed tool takes a virtualization-obfuscated executable as input, extracts its execution trace, and provides the operational process of each handler along with the pre- and post-execution state as symbolic execution expressions. 
Security analysts can leverage these results to reduce analysis time and readily identify critical handler information in malware. 
Steps (1) and (2) are responsible for acquiring semantic anchors, while steps (3) and (4) perform semantic extraction and verification.

(1) 
Trace Extractor generates an execution trace of the target executable using a Dynamic Binary Instrumentation (DBI) tool such as Intel Pin. 
The trace records the following information for each executed instruction: trace sequence number, instruction address, assembly representation, hexadecimal encoding, memory read/write flags along with the accessed address and size, register value changes before and after execution, and the current value of the rsp register. 
This trace information is then used to identify the virtualization-obfuscated region.

(2) 
Handler Detector identifies virtual instruction handlers. 
First, based on the observation that the result of any virtual instruction handler must be stored in the virtual context, it computes the address range of the virtual context analyzed in Section 2.B relative to rsp, and identifies store instructions targeting memory within that range as result-storing instructions. 
It then traces backwards from each result-storing instruction to the most recently preceding indirect jump instruction, treating the resulting slice as a single handler block for handler identification.

\begin{figure}[t]
    \centering
    \includegraphics[width=0.8\linewidth]{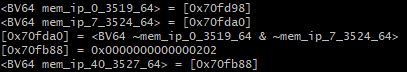} 
    \caption{Analysis Result of the Virtual Instruction Handler (V\_NOR) by VMPredator} 
    \label{fig:f3}
\end{figure}

(3) 
Semantic Extractor applies symbolic execution to each handler block to generate expressions representing the handler's operational process. 
By extracting only the expressions that compute values stored in the virtual context, it automatically eliminates garbage instructions and decoding routines inserted within the handler. 
As a result, applying this process to all executed handler blocks yields a symbolic representation of the operations ultimately performed by the program under analysis. 
The expressions are subsequently simplified through SMT solver-based optimization. 
The resulting expressions describe either a handler block reading a value from a specific memory address and storing it in a physical register, or a handler block storing its computed result to a specific memory address. 
Figure \ref{fig:f3} shows the Handler Expression output by the Semantic Extractor for the V\_NOR handler. 
The analysis result reveals that the handler reads two values from the virtual stack, performs a NOR operation, and stores the result back onto the virtual stack.

\begin{figure}[t]
    \centering
    \includegraphics[width=0.8\linewidth]{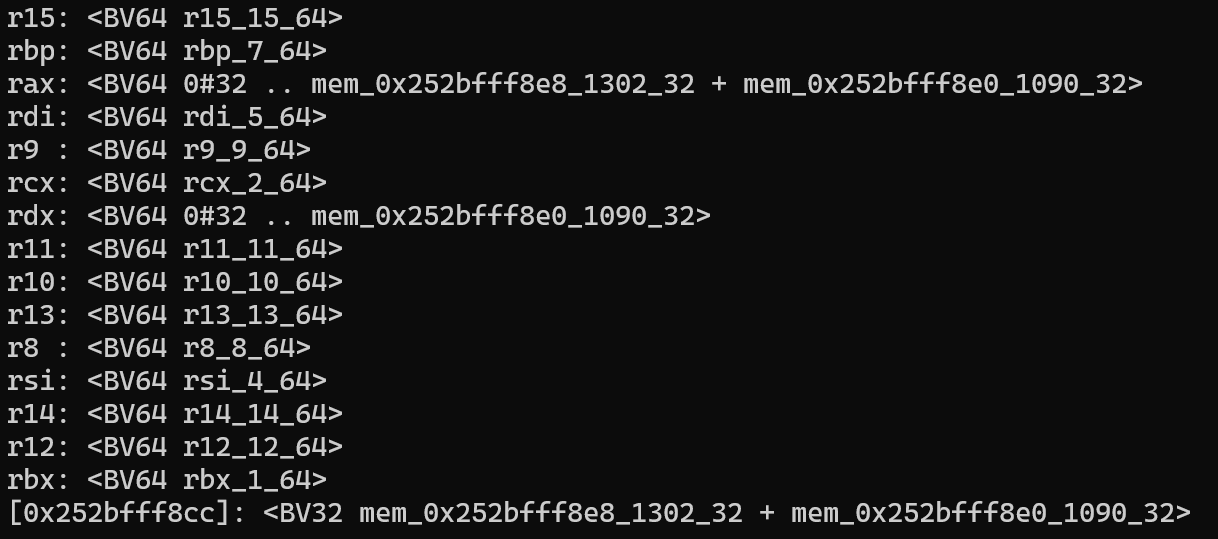} 
    \caption{Output of the Expression Emulator in VMPredator} 
    \label{fig:f4}
\end{figure}

(4) 
Expression Emulator executes the Handler Expressions produced by the Semantic Extractor and outputs State Expressions -- symbolic expressions representing the values stored in registers and non-virtual-context memory addresses after the virtual machine has finished execution. 
Figure \ref{fig:f4} shows an example of the State Expression output by the Expression Emulator. 
The original program implements r = x + y, which was obfuscated and then analyzed; a total of 83 virtual instruction handlers were detected. 
Among these, the rax register holds the result of the expression x + y, and memory address 252bfff8cc corresponds to the variable r. 
In this way, expressions stored in each register and non-virtual-context memory region are extracted and represented as symbolic execution expressions, making them readily available for subsequent analysis.

\section{Implementation and Evaluation}
VMPredator was implemented using the trace extraction functionality of Intel Pin 4.0\cite{b2} and angr 9.2\cite{b6}. 
The obfuscation tool used in the experiments was VMProtect 3.5.1 Ultimate, and the test code consisted of C and assembly language programs generated with the assistance of an LLM. 
The following research questions were formulated to guide the experimental evaluation:

RQ1: Are the analysis results produced by VMPredator accurate?

RQ2: Does VMPredator offer improved performance compared to existing virtualization deobfuscation tools?

\begin{table}[b]
\centering
\caption{Performance of the Proposed VMPredator} 
\label{table:t1}
\begin{tabular}{c|ccccc}
\hline
\begin{tabular}[c]{@{}c@{}}original \\ instructions\end{tabular}            & \multicolumn{5}{c}{}                                                                                                                                                                                                                                                                             \\ \hline
\multirow{5}{*}{\begin{tabular}[c]{@{}c@{}}3 \\ instructions\end{tabular}}  & test cases                                                                         & \multicolumn{2}{c}{exact  correct cases}                                                                                   & \multicolumn{2}{c}{wrong  cases}                                               \\
                                                                            & 1,141                                                                              & \multicolumn{2}{c}{1,141}                                                                                                  & \multicolumn{2}{c}{0}                                                          \\
                                                                            & \multirow{2}{*}{\begin{tabular}[c]{@{}c@{}}obfuscated \\ trace lines\end{tabular}} & \multirow{2}{*}{\begin{tabular}[c]{@{}c@{}}result \\ lines\end{tabular}} & \multicolumn{2}{c}{\multirow{2}{*}{\begin{tabular}[c]{@{}c@{}}detected \\ handlers\end{tabular}}} & \multirow{2}{*}{time (sec)}  \\
                                                                            &                                                                                    &                                                                          & \multicolumn{2}{c}{}                                                                              &                              \\
                                                                            & 5,466                                                                              & 832                                                                      & \multicolumn{2}{c}{88}                                                                            & 90                           \\ \hline
\multirow{5}{*}{\begin{tabular}[c]{@{}c@{}}20 \\ instructions\end{tabular}} & test cases                                                                         & \multicolumn{2}{c}{exact   correct cases}                                                                                  & \multicolumn{2}{c}{wrong  cases}                                               \\
                                                                            & 1,051                                                                              & \multicolumn{2}{c}{1,048}                                                                                                  & \multicolumn{2}{c}{3}                                                          \\
                                                                            & \multirow{2}{*}{\begin{tabular}[c]{@{}c@{}}obfuscated \\ trace lines\end{tabular}} & \multirow{2}{*}{\begin{tabular}[c]{@{}c@{}}result \\ lines\end{tabular}} & \multicolumn{2}{c}{\multirow{2}{*}{\begin{tabular}[c]{@{}c@{}}detected \\ handlers\end{tabular}}} & \multirow{2}{*}{time (sec)}  \\
                                                                            &                                                                                    &                                                                          & \multicolumn{2}{c}{}                                                                              &                              \\
                                                                            & 15,262                                                                             & 2,461                                                                    & \multicolumn{2}{c}{259}                                                                           & 193.9                        \\ \hline
\end{tabular}
\end{table}

RQ1: To evaluate the accuracy of VMPredator, we assessed whether the simplification results match the behavior of the original binary. 
We used Claripy from angr to check whether the State Expressions extracted from the obfuscated trace by the proposed tool are logically equivalent to the expressions stored in each register after executing the original binary. 
For incorrect cases, we measured the proportion of correct registers out of all registers for each case. 
Table \ref{table:t1} presents the performance of VMPredator. 
For files consisting of 3 instructions, all results were confirmed to be equivalent. 
For data consisting of 20 instructions, 3 incorrect results were observed. 
Analysis of the root cause revealed that in some handlers performing complex computations, 
the AST depth of the extracted Claripy expressions exceeded 6, causing values to be substituted with constants and the computation process to be lost.
The experimental results show that analysis was completed within the time limit for all test data. 
The average analysis time for 20-instruction data was 194 seconds. 
Furthermore, while the obfuscated traces exhibited approximately 950 to 1,800 times greater complexity compared to the original, the information extracted by VMPredator was reduced by approximately 85\% relative to the obfuscated traces.

\begin{table}[t]
\centering
\caption{Comparison of VMPredator and \cite{b4}} 
\label{table:t2}
\setlength{\tabcolsep}{2pt}
\begin{tabular}{c|ccccc}
\hline
\cellcolor[HTML]{FFFFFF}                         & \cellcolor[HTML]{FFFFFF}                                                                       & \cellcolor[HTML]{FFFFFF}                                                                      & \cellcolor[HTML]{FFFFFF}                                                                                   & \cellcolor[HTML]{FFFFFF}                                                                                          & \cellcolor[HTML]{FFFFFF}                                                                                           \\
\multirow{-2}{*}{\cellcolor[HTML]{FFFFFF}Sample} & \multirow{-2}{*}{\cellcolor[HTML]{FFFFFF}\begin{tabular}[c]{@{}c@{}}Code \\ Type\end{tabular}} & \multirow{-2}{*}{\cellcolor[HTML]{FFFFFF}\begin{tabular}[c]{@{}c@{}}Arg \\ Type\end{tabular}} & \multirow{-2}{*}{\cellcolor[HTML]{FFFFFF}\begin{tabular}[c]{@{}c@{}}VMPredator \\ (proposed)\end{tabular}} & \multirow{-2}{*}{\cellcolor[HTML]{FFFFFF}\begin{tabular}[c]{@{}c@{}}\cite{b4} with \\ Provided \\ Trace (a)\end{tabular}} & \multirow{-2}{*}{\cellcolor[HTML]{FFFFFF}\begin{tabular}[c]{@{}c@{}}\cite{b4} with \\ Generated \\ Trace (b)\end{tabular}} \\ \hline
Sample1                                          & \begin{tabular}[c]{@{}c@{}}Simple \\ Arithmetic\end{tabular}                                   & Int                                                                                           & Correct                                                                                                    & Correct                                                                                                           & Correct                                                                                                            \\ \hline
Sample2                                          & \begin{tabular}[c]{@{}c@{}}Simple \\ Boolean\end{tabular}                                      & Int                                                                                           & Correct                                                                                                    & Correct                                                                                                           & Correct                                                                                                            \\ \hline
Sample3                                          & \begin{tabular}[c]{@{}c@{}}Complex \\ MBA \\ Expression\end{tabular}                           & Char                                                                                          & Correct                                                                                                    & Correct                                                                                                           & Correct                                                                                                            \\ \hline
Sample4                                          & Simple if                                                                                      & Int                                                                                           & Correct                                                                                                    & Correct                                                                                                           & Correct                                                                                                            \\ \hline
Sample5                                          & \begin{tabular}[c]{@{}c@{}}Simple if with \\ Arithmetic\end{tabular}                           & Int                                                                                           & \textbf{Correct}                                                                                           & Correct                                                                                                           & \textbf{Timeout}                                                                                                   \\ \hline
Sample6                                          & \begin{tabular}[c]{@{}c@{}}Simple if with \\ arithmetic\end{tabular}                           & Unsigned                                                                                      & \textbf{Correct}                                                                                           & Correct                                                                                                           & \textbf{Timeout}                                                                                                   \\ \hline
\end{tabular}
\end{table}

RQ2: Most previously proposed methods and tools for virtualization obfuscation analysis have not been publicly released and could not be reproduced\cite{b4}. 
In this paper, we compare against \cite{b4}, which is regarded as the state-of-the-art among publicly available automated tools. 
The comparison dataset consists of 6 samples provided by \cite{b4}, and the composition of the experimental dataset is shown in Table ?. 
The dataset spans a diverse range of types, from simple arithmetic operations to complex expressions and conditional branches. 
The dataset is divided into two categories: (a) traces of binaries obfuscated with an unknown version of VMProtect, directly provided by \cite{b4}; and (b) traces extracted by this study from binaries obfuscated using the threaded variant of VMProtect. 
The comparison results show that VMPredator completed analysis on all datasets.
In contrast, the tool from \cite{b4} is designed for the decode-dispatch loop approach, and as a result, it failed to complete analysis within the time limit on a subset of the (b) data obfuscated using the threaded approach.

The two experiments above confirm that the proposed tool can effectively simplify obfuscated programs while preserving the semantics of the original pre-obfuscation code, and that it is capable of analyzing threaded virtualization obfuscation, which existing related work cannot handle.
\section{Conclusion and Future Work}
This study proposed VMPredator, a tool for automatically extracting the operational semantics of virtualization-obfuscated programs. 
By defining semantic anchors, the tool is applicable to diverse virtualization structures. 
Experiments confirmed complete semantic equivalence with the original for small-scale programs, and comparison with related work demonstrated superior analysis capability against threaded obfuscation. 
Future work will focus on expanding the accuracy and applicability of the tool through conditional branch analysis for control flow recovery, improvement of the AST depth-based decoding identification mechanism, and automatic detection of obfuscated regions.

\end{document}